\documentclass{dsj}

\usepackage{fancyhdr}
\usepackage{lipsum}
\usepackage{graphicx}
\usepackage{caption}
\usepackage{array}

\usepackage{longtable}
\usepackage{adjustbox}
\usepackage{parskip}
\usepackage{multirow}
\usepackage{fancyhdr}
\usepackage[labelfont=bf]{caption}
\begin{document}
\pagestyle{fancy}

\submitted{4 November 2014}

\title{A Conceptual Framework for Designing Interactive Human-Centred Building Spaces to Enhance User Experience in Specific-Purpose Buildings }

\author[a]{Roofigari-Esfahan, N.\email{nazila@vt.edu}}
\author[b]{Morshedzadeh, E}
\author[c]{Dongre, P}

\address[a]{Department of Building Construction, Virginia Tech, Blacksburg, VA 24061 USA}
\address[b]{ Industrial Design Program, Gerald Hines College of Architecture and Design, University of Houston, Houston, TX 77470 USA, }
\address[c]{Department of Computer Science, Virginia Tech, Blacksburg, VA 24061 USA}

\maketitle

\begin{abstract}
Human/User interaction with buildings are mostly restricted to interacting with building automation systems through user-interfaces that mainly aim to improve energy efficiency of buildings and ensure comfort of occupants. This research builds on the existing theories of Human-Building Interaction (HBI) and proposes a novel conceptual framework for HBI that combines the concepts of Human-Computer Interaction (HCI) and Ambient Intelligence (AmI). The proposed framework aims to study the needs of occupants in specific-purpose buildings, which is currently undermined. Specifically, we explore the application of the proposed HBI framework to improve the learning experience of students in academic buildings. Focus groups and semi-structured interviews were conducted among students who are considered primary occupants of Goodwin Hall, a flagship smart engineering building at Virginia Tech. Qualitative coding and concept mapping were used to analyze the qualitative data and determine the impact of occupant-specific needs on the learning experience of students in academic buildings. The occupant-specific problem that was found to have the highest direct impact on learning experience was “finding study space” and highest indirect impact was “Indoor Environment Quality (IEQ)”. We discuss new ideas for designing Intelligent User Interfaces (IUI), e.g. Augmented Reality (AR), increase the perceivable affordances for building occupants and considering a context-aware ubiquitous analytics-based strategy to provide services that are tailored to address the identified needs.

\subsubsection{\upshape Keywords}
Ambient intelligence; Sensor networks; Human Building Interaction; Occupant engagement; User experience

\end{abstract}

\footnotetext{\footnotesize Abbreviations:BAS- Building automation system; building management staffs (BMS); HBI- Human building interaction; HCI- Human computer interaction; AmI- Ambient intelligence; IEQ-Indoor environment quality; IUI- Intelligent user interface; ML- Machine learning; NLP- Natural language processing; ABM- Agent-based modeling; RTLS- Real-time location sensing; VOA- Voice operated assistant; DHH- Data hungry home; IR- Infrared; RFID- Radio frequency identification devices; BLE- Bluetooth Low Energy; UWB- Ultra-Wide Band; RH- Relative humidity; LA- Learning analytics}\hfill

\section{}
We live in a world that involves various levels of interactions, with each other (living entities) and the spaces in which we live. Interactions that we can code using data. We experience this world by navigating through the spaces created by the architects and by the systems designed that define the utility of a particular space. Thus, data plays a vital role in understanding the impact of any change or any system that is currently in place. This is where HCI (Human Computer Interaction) methods prove to be useful. They act as a skeleton on which HBI (Human Building Interaction) attaches itself as a coherent entity. HBI grows with the potential of HCI, and they come together in this research to provide a framework for studying occupant/user problems.

\subsection{1. Introduction}
With the advancement in technology, buildings are transforming into smart environments. Smart buildings are equipped with building automation systems (BAS) that aim to automate various building operations such as Heating, Ventilation, and Air Conditioning (HVAC), energy consumption, lighting, etc. These systems operate autonomously, and human involvement is restricted to building user-interfaces (UIs). The current BASs are mostly energy-focused \citep{bib1} and aim to enable building management staffs (BMS) to access the building systems through controllers and interaction points. A typical BAS is composed of sensors to collect data regarding building temperature, humidity, occupancy, etc. sensor networks that connect the building sensor nodes, control systems that take data from the sensor network and make decisions for the building; output devices (actuators) to execute the decisions made by the control system, and UIs for building managers and occupants to interact with the BAS. 
With the objective of increasing human (BMS and general occupants) interaction with BAS and engagement in the design, operation and maintenance of buildings, novel research area called “Human-Building Interaction (HBI)” has emerged \citep{bib2}. Human-Building Interaction (HBI) is a convergent field that represents the growing complexities of the dynamic interplay between human experience and intelligence within built environments \citep{bib3}. Some researchers explain HBI by comparing it with Human-Computer Interaction (HCI) which is a field of study that explores interactions between humans and computer systems to design user-friendly UIs \citep{bib4}. The HCI-based approach defines HBI as the application or channel of HCI in the domains of architecture and urban design, to enable efficient occupant interactions with the built environment focusing on human values, needs, and priorities \citep{bib5}. HBI and the bilateral impacts that building and its occupant’s behavior have on each other has been a point of interest in much recent research for increasing the energy efficiency \citep{bib6},\citep{bib7}, investigating occupant comfort \citep{bib8} and interactions \citep{bib9},\citep{bib10}. Alavi et al. \citep{bib5} divided the scope of HBI research into three categories namely “People/User”, “Built Environment/Building”, and “Computing”. Their proposed diagram has three interrelated dimensions of “Physical”, “Spatial”, and “Social” which describe various HBI research directions. Using three concentric circles (Fig. 1).

\begin{figure}[ht]
    \centering
    \captionsetup{justification=centering}
\includegraphics[width=5cm]{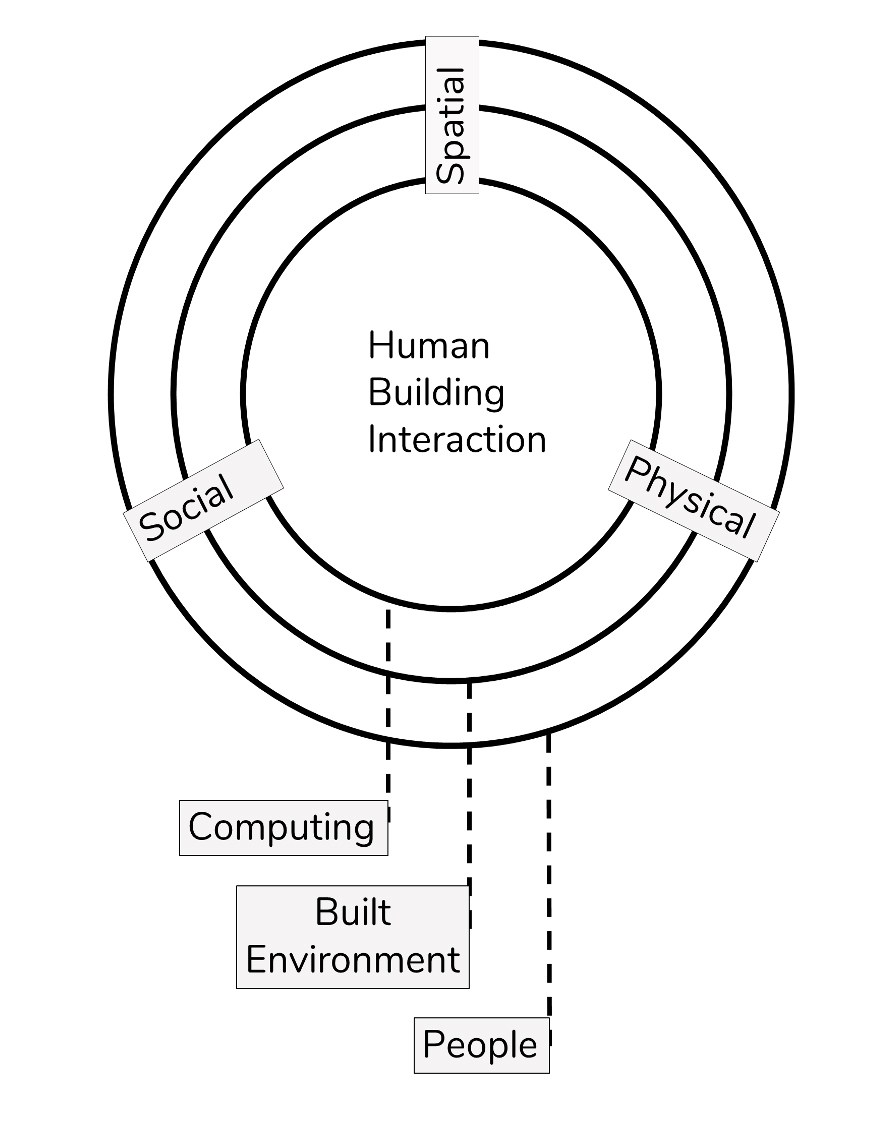}
\caption{\footnotesize \label{fig1}HBI Scope-Dimension Diagram [5]}
\end{figure}

The “Physical” represents the environmental conditions such as temperature and humidity; “Spatial” represents attributes such as building space utilization, and “Social” represents collaboration and negotiation between building occupants such as indoor traffic management. These aspects are corelated to each other and can be combined to propose new research and design directions. For instance, a combined example of “Spatial” and “Social” can be demonstrated by smart collaborative workspaces \citep{bib5}.
Other researchers explain HBI by comparing it with Ambient Intelligence (AmI) which is the next step for Artificial Intelligence (AI). Contrary to the HCI-based definition that focuses on occupant participation, needs, and values, the AmI-based definition of HBI is concerned with increasing the intelligence of the building environments using AI. 
\begin{figure}[ht]
    \centering
    \captionsetup{justification=centering}
\includegraphics[width=5cm]{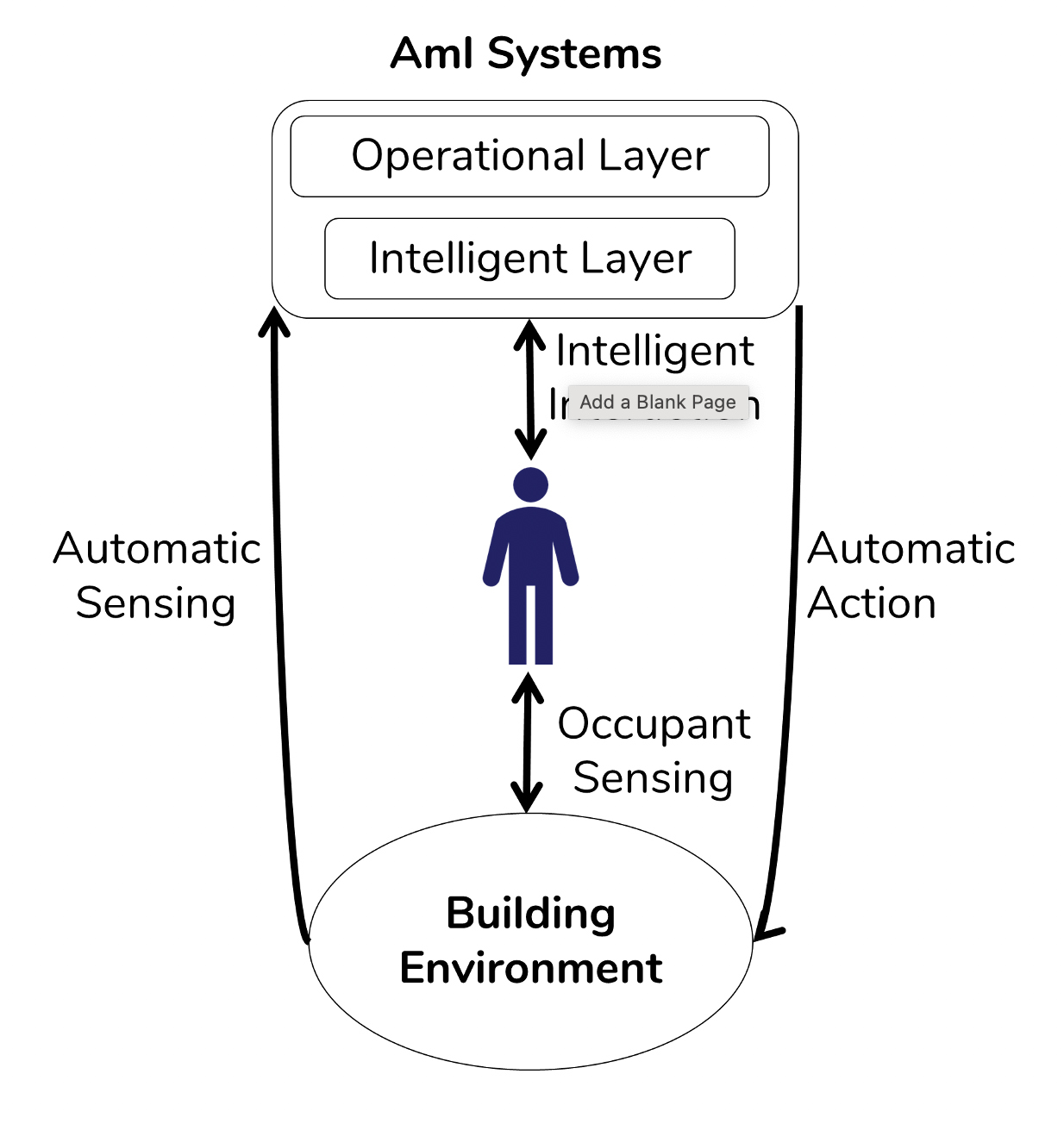}
\caption{\footnotesize \label{fig2}Ambient Intelligence (AmI) Framework}
\end{figure}
It makes use of ubiquitous computing devices in buildings to add various capabilities such as awareness of occupant needs, intelligent interaction with occupants, forecasting occupant behavior, and taking necessary actions \citep{bib11}. AmI systems comprise of two layers, namely the operational layer and the intelligent layer. The operational layer in AmI systems consists of operating systems, databases, ubiquitous computing, etc. The added benefit of building AmI systems over BAS is the presence of an intelligent layer that uses computational methods such as Machine Learning (ML), Natural Language Processing (NLP), etc. for performing predictive analytics \citep{bib12},\citep{bib13}. The data for computation in AmI systems is collected from occupants and building environments using sensors. AmI systems automatically take necessary actions by using output devices such as actuators and robots \citep{bib14},\citep{bib15}. An illustration of an AmI framework is shown in Figure 2.
HBI applications mostly focus on saving energy, increasing occupant’s comfort, and enhancing occupant’s safety by facilitating interaction between occupants and buildings. Wang \& Heydarian \citep{bib16} proposed an approach to collect and analyze psychological and environmental data to build occupant behavior models and pair them with targeted interventions to increase energy efficiency. Jia et. al. \citep{bib2} used agent-based modelling (ABM) for developing occupant behavior models from the data collected using various sensors and surveys to improve the accuracy of energy estimation. Abraham et. al. \citep{bib17} used ML methods to train the occupant behavior-related energy utilization data and predict the energy consumption pattern of a building. A socio-technical energy management system by the name “BizWatts” was developed by Gulbinas et. al. \citep{bib18} to save energy by providing real-time, appliance-level power management and socially contextualized energy consumption feedback to the occupants. In continued research, they used BizWatts to understand the impact of organizational occupant behavior on energy savings \citep{bib19}. Occupant energy consumption data was collected using smart plugs and the comparison showed to the users on a web-based interface. However, users cannot always be trusted to maintain positive energy behavior. Ahmadi-Karvigh et. al. \citep{bib20} developed algorithms to recognize user actions and user activities at peak and non-peak hours to estimate the energy wasted inside a building. 
Besides reducing energy consumption, users’ comfort also should be considered to maximize building efficiency \citep{bib9}. Nguyen et. al. \citep{bib21} developed a system that deals with advanced load management strategies and real-time wireless communication techniques to reduce the peak consumption while maintaining thermal comfort. Alavi et al. \citep{bib22} proposes a schematic model of comfort and demonstrates an interactive tool called “Comfort Box” that collects subjective feedback from occupants about the perception of comfort in buildings. Jazizadeh et. al. used building data obtained from sensors and occupant data obtained from wearable devices to develop personalized thermal comfort prediction models \citep{bib23},\citep{bib24},\citep{bib25}. Similar models were also used by Li et. al. to determine optimum HVAC control strategies for buildings \citep{bib26}. In another stream of HBI research, building and occupant data were used to make best use of building spaces. Verma et. al. used sensing and participatory data to understand how occupants use spaces in a building, aiming at optimizing building space utilization \citep{bib26}. “Twitter Bots”, autonomous tweeting robotic agents, were used to engage occupants in a process of providing everyday feedback about space use \citep{bib27}. Understanding occupants’ behavior in building emergencies can help in the designing safer buildings \citep{bib28},\citep{bib29}. To ensure occupant safety inside the building, Cheng et. al. used building sensor data and a smartphone application to aid the occupants in case of fire emergencies \citep{bib28}. Chen et. al. utilized sensor data, guiding firefighters to quickly locate the fire inside a building, using LED light indicators placed at various locations in the building \citep{bib30}. 
In reviewing different HBI theories and applications, gaps in considering real-time bidirectional, and multi-modal interactions between occupants and buildings were found. Also impact of HBI in building with specific functions such as academic buildings and their impact on the goals of their particular occupants needs more exploration. This study proposes a vision for developing a Human Building Interaction (HBI) framework that combines AmI and HCI and focuses on developing a conceptual HBI framework for educational buildings, aiming at improving students’ experience in these buildings. 

\subsection{2. Research Objectives}
In order to address the aforementioned gaps, the objective of this research is to, first, develop a holistic HBI framework that combines HCI and AmI. The goal of the proposed framework is to enable bidirectional interaction between buildings and their occupants, to create more intelligent and interactive building environments. We then investigate the application of the proposed framework in enhancing the learning experience of students (as primary occupants) in academic buildings. We hypothesize that addressing occupant-specific problems in academic buildings can have a positive impact on the learning experience of students, as the primary occupants. To this end, we first investigate the occupant-specific problems that can diminish the learning experience of students in academic buildings. We then use the proposed HBI system as an intervention to mitigate the occupant-specific problems and enhance the learning experience of students in academic buildings. 

\subsection{3. Proposed Human Building Interaction (HBI) Scope-Dimension Diagram }
The proposed extended HBI scope-dimensions diagram is shown in Figure 3and was developed following the steps described here. Firstly, we expand the scope of the HBI research that is represented by “People,” “Built Environment,” and “Computing”. The scope “People” denotes the involvement of occupants in HBI and can be collected through participatory involvement of occupants or use of sensors. In participatory involvement, occupants are contributing via surveys, questionnaires, etc. (active engagement), whereas sensory involvement includes occupant data collection via sensing technologies (passive engagement). The scope “Built Environment” denotes the level of interaction that occupants have with the building. These interactions can be basic, e.g., opening and closing of a door, or smart, e.g., use of interfaces to collect data about perceivable comfort levels of occupants and regulating the Indoor Environment Quality (IEQ) parameters inside the building. The scope “Computing” denotes the level of analysis performed on the collected data from participation and sensing and can be operational or intelligent. Operational computing is supported by operating systems, communications, databases, ubiquitous computing, etc., while intelligent computing is supported by ML, NLP, Computer Vision, etc.

\begin{figure}[ht]
    \centering
    \captionsetup{justification=centering}
\includegraphics[width= 12cm]{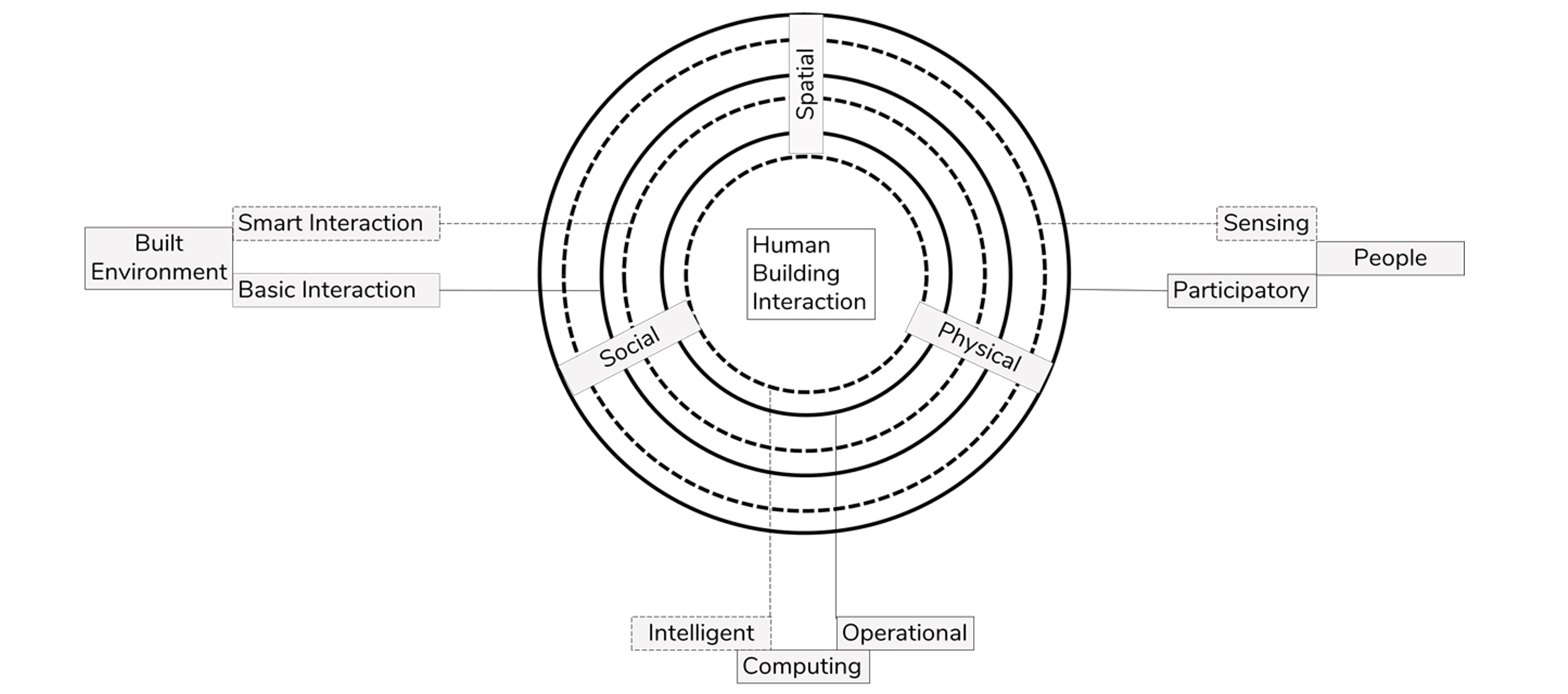 }
\caption{\footnotesize \label{fig3}Proposed HBI Scope-Dimension Diagram}
\end{figure}

Existing HBI literature, when mapped on the proposed diagram, revealed some interesting insights. The mapped HBI research is shown as oval shapes on the proposed scope-dimension diagram shown in Figure 4. As previously mentioned, common HBI applications include improving energy savings, occupant comfort, and occupant safety this can be observed on the Figure 4. that energy-driven applications (to increase energy efficiency) lean towards physical aspect of physical-social interactions and are more widely studied to cover a range of interactions and applications including smart interactions, operational computing, and intelligent computing. The comfort driven studies lean more towards social aspect of social-physical interactions representing those developing human-behavior models to control energy consumption and thermal comfort. Safety driven studies, e.g., indoor real-time location sensing (RTLS) for emergency navigation, covers the physical-spatial dimension. The studies in this category have a limited scope, using sensory data received from indoor RTLS, and not including participants subjective feedback through participatory data and intelligent computing.  

\begin{figure}[ht]
    \centering
    \captionsetup{justification=centering}
\includegraphics[width= 12cm]{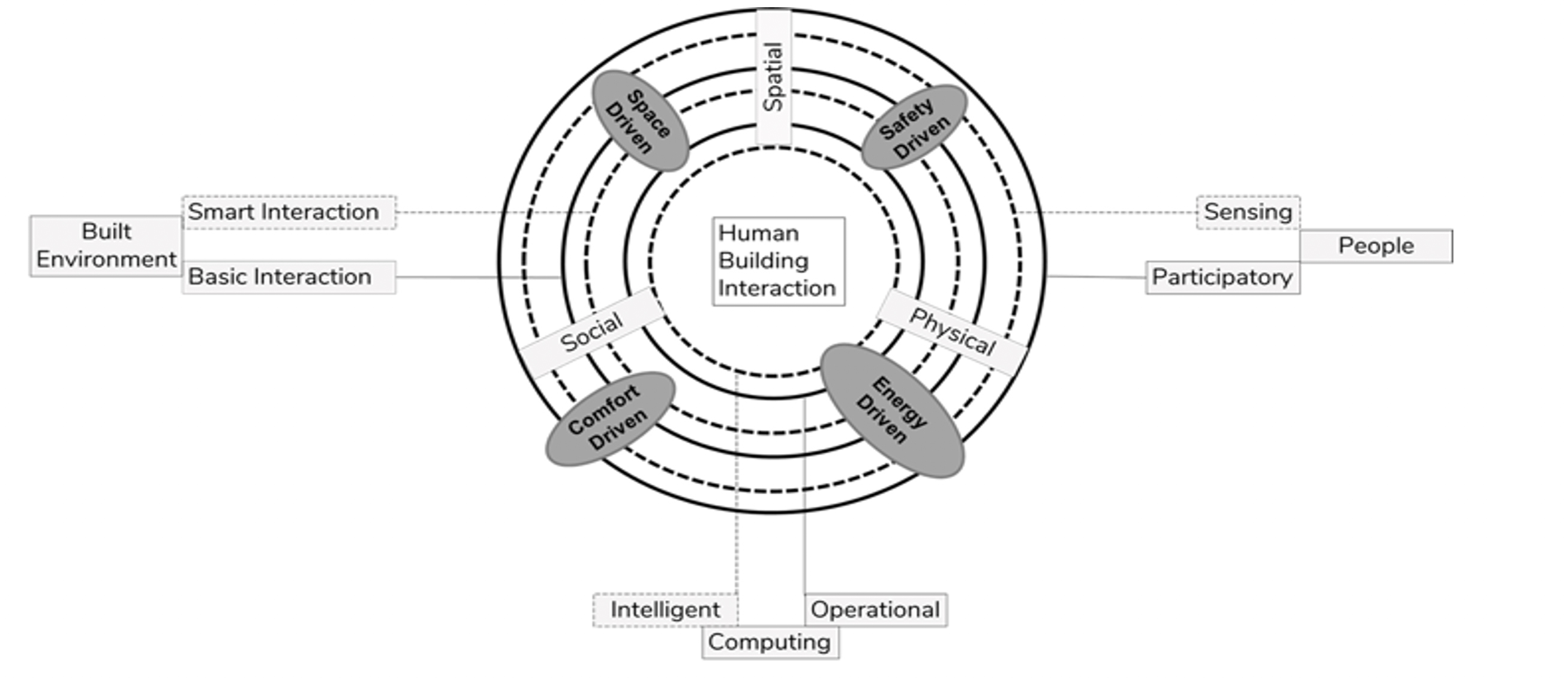}
\caption{\footnotesize\label{fig4}Existing HBI Applications mapped on the proposed Scope-Dimension Diagram}
\end{figure}

Finally space driven research, e.g., use of RTLS to increase the spatial comfort of occupants, denotes the oval shape in the spatial-social dimension. Mapping existing application on the proposed scope-dimension diagram can also be used to identify future research directions for HBI applications. Additionally, as previously stated, there is a need for a holistic framework for HBI to ameliorate the bi-directional and intelligent interaction between buildings and their occupants. Using the proposed scope-dimension diagram, a holistic HBI framework is developed that augments current HBI frameworks with combined benefits of HCI and AmI to enable bi-directional and intelligent interaction between buildings and their occupants. The Proposed framework is then applied to a proof-of-concept study to explore the potential of proposed approach in improving occupants’ experience in specific-purpose buildings, including student learning in academic buildings. 

 \subsection{4. Proposed Extended Human Building Interaction (HBI) Framework}
For a long time, HCI and AI were considered divergent fields of study because HCI encourages human involvement to make UIs more user-friendly whereas AI aims to construct human-like intelligence by using mathematical models \citep{bib31}. However, researchers from the HCI community and AI community are now coming together to improve the interaction between humans and AI systems through designing intelligent user interfaces (IUIs) \citep{bib32}. Voice and text operated assistants (VOAs) such as Apple’s Siri, Google Assistant, Amazon’s Alexa and ChatGPT are some examples of IUIs. Similarly, the HCI and AmI-based definitions of HBI can be combined. In this research, we propose a novel vision for an HBI Framework that makes use of this integration, aimed at creating IUIs for buildings as shown in Figure 5.

\begin{figure}[ht]
    \centering
    \captionsetup{justification=centering}
\includegraphics[width= 8cm]{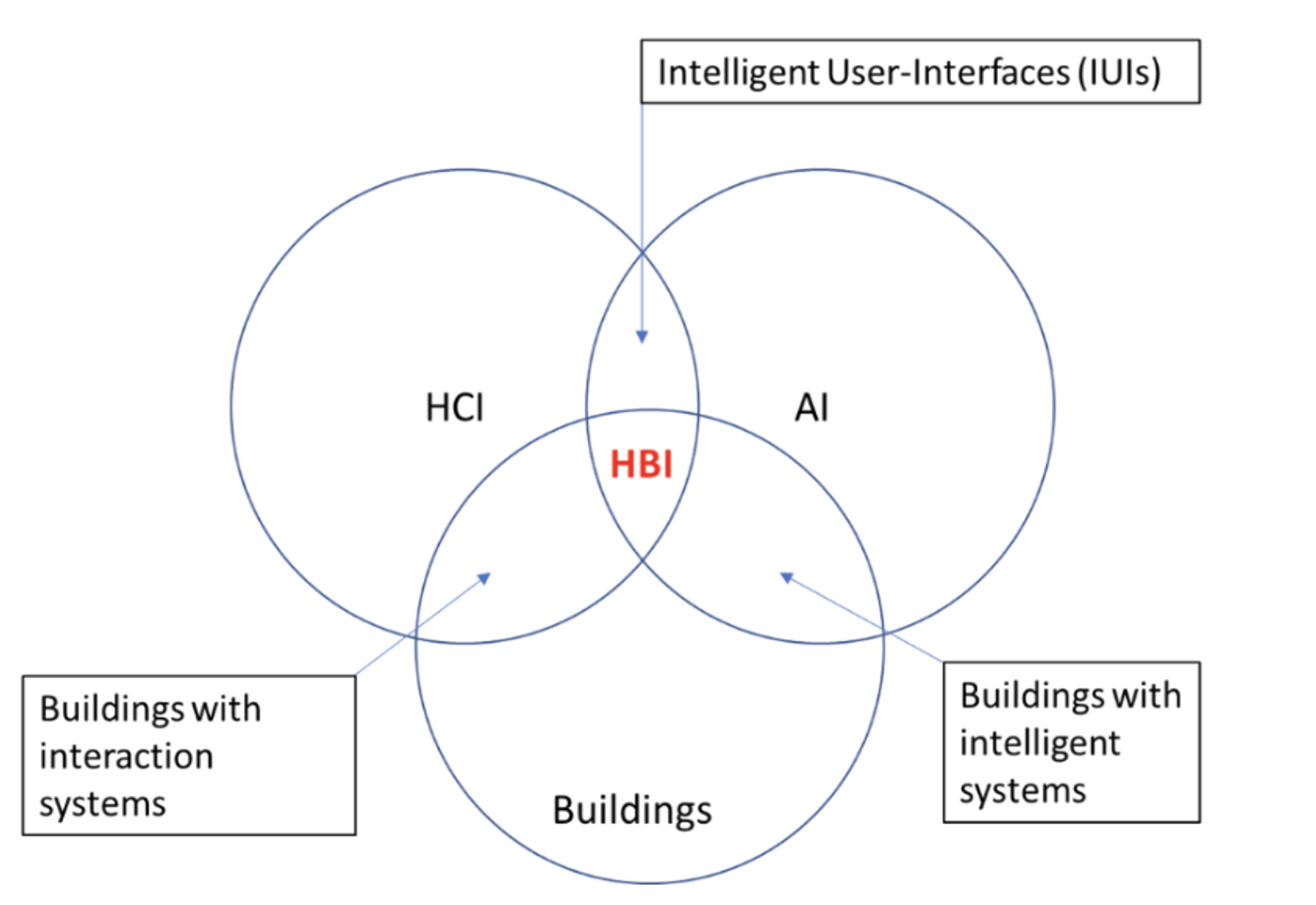 }
\caption{\footnotesize\label{fig5}Proposed approach to Intelligent User Interfaces (IUIs) design for HBI}
\end{figure}

The overlap between HCI and buildings represent buildings with interfaces to facilitate interaction between occupants and buildings. The overlap between AI and buildings represents buildings that have AI-based systems to intelligently control building operations. Our vision of HBI lies at the intersection of HCI, AI and buildings, and focuses on improving the interactions between occupants and building through designing IUIs for buildings, aiming to facilitate real-time, bidirectional, and multi-modal interactions between the occupants and buildings. 
Figure 6 illustrates a model of the proposed HBI framework that is comprised of an intelligent layer for predictive analytics and IUI for facilitating real-time, bidirectional, and multi-modal interactions between the occupants and buildings. In this framework, data required for computation is collected from occupants through sensing and participation and from buildings through real-time sensing and historical data in the building database, as opposed to conventional sensing in AmI systems. Participatory data ensure the active involvement of occupants in the HBI system. 

\begin{figure}[ht]
    \centering
    \captionsetup{justification=centering}
\includegraphics[width= 8cm]{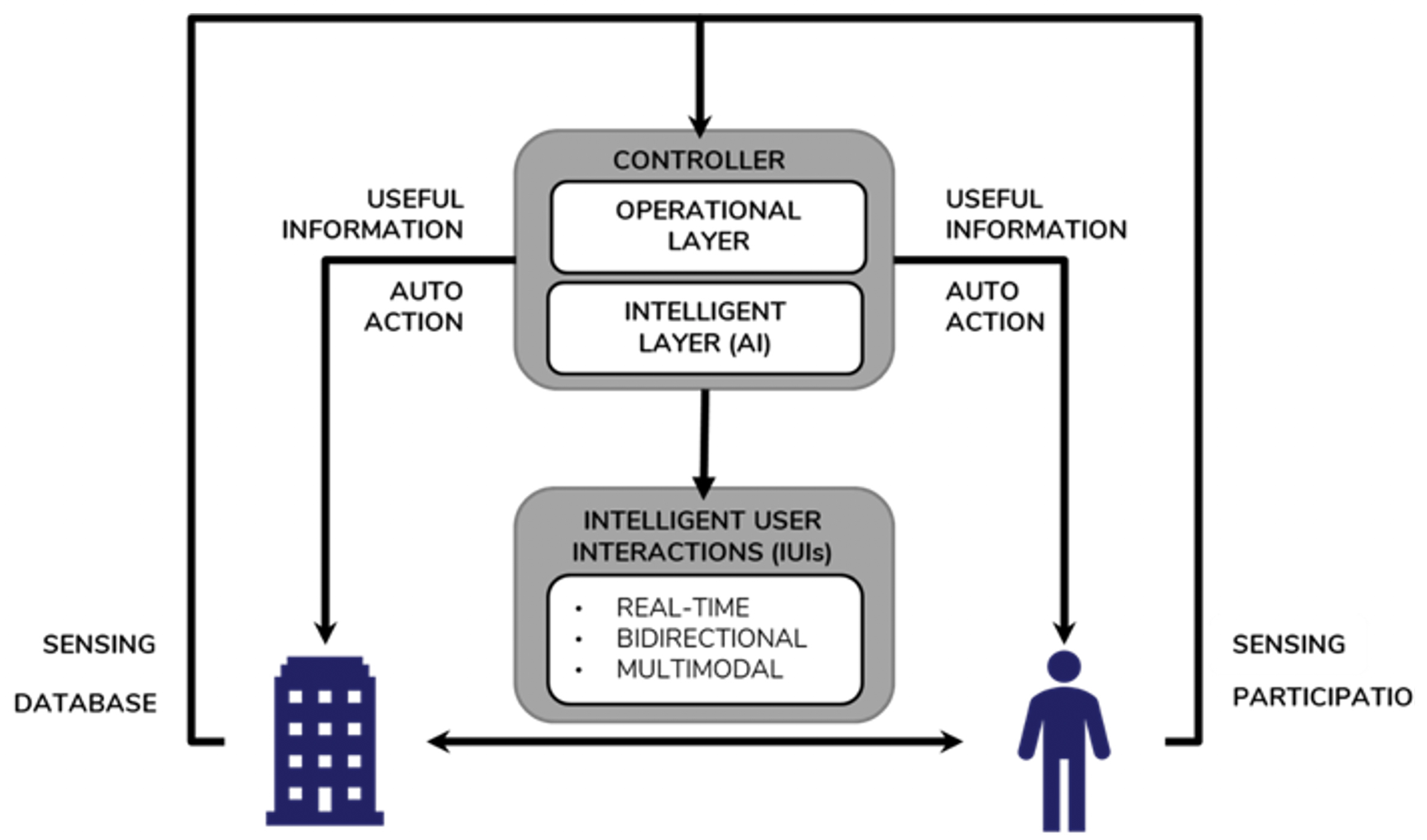 }
\caption{\footnotesize \label{fig6}Proposed Human Building Interaction (HBI) Framework }
\end{figure}

On the other hand, the building database contains all essential data about the building that cannot be measured using sensing techniques, including as-built plans. Similar to the AmI systems, the controller in the proposed HBI framework has an intelligent layer that uses AI to forecast occupant behavior and building requirements. 
The decisions made by the controller are translated into useful information and automatic actions. These automatic actions benefit both the occupants and buildings, by providing context-aware information or guidance to occupants or actuating specific sensors in the building. The framework then conveys information through multi-modal interactions such as audio, touch, gestures, smell, etc. that consequently increases the efficiency of interaction between occupants and buildings \citep{bib32}. 

\subsection{5. Application of the proposed HBI Framework for Specific-Purpose Buildings}
The proposed HBI framework takes a human-centered approach in establishing real-time, bidirectional, and multi-modal interactions between occupants and buildings that are tailored to their needs. The following steps map the approach to create the interactions: 1) identify building type and its primary occupants, 2) identify primary occupants’ specific problems and preferences, 3) identify the data types and respective data acquisition methods. (This data is required to address the identified occupant problems), 4) collect and analyze data and extract the required knowledge based on the application, and lastly 5) develop HBI Intelligent User Interfaces (IUIs) to convey the acquired knowledge to the occupants. In this section, the application of the proposed framework to enhance the learning experience of primary occupants in academic buildings (students) along with the case study that was carried out in a building on Virginia Tech Campus is presented. To this end, first, the occupant-specific problems were identified and their impact on the learning experience of students was evaluated. The existing BAS in the selected building (Goodwin Hall) was also assessed. Finally, potential technologies to complement the current BAS to achieve the added benefits of the proposed HBI framework to improve the learning experience of students were evaluated and suggestions were provided. The research steps taken are shown in Figure 7.
\begin{figure}[ht]
    \centering
    \captionsetup{justification=centering}
\includegraphics[width= 10cm]{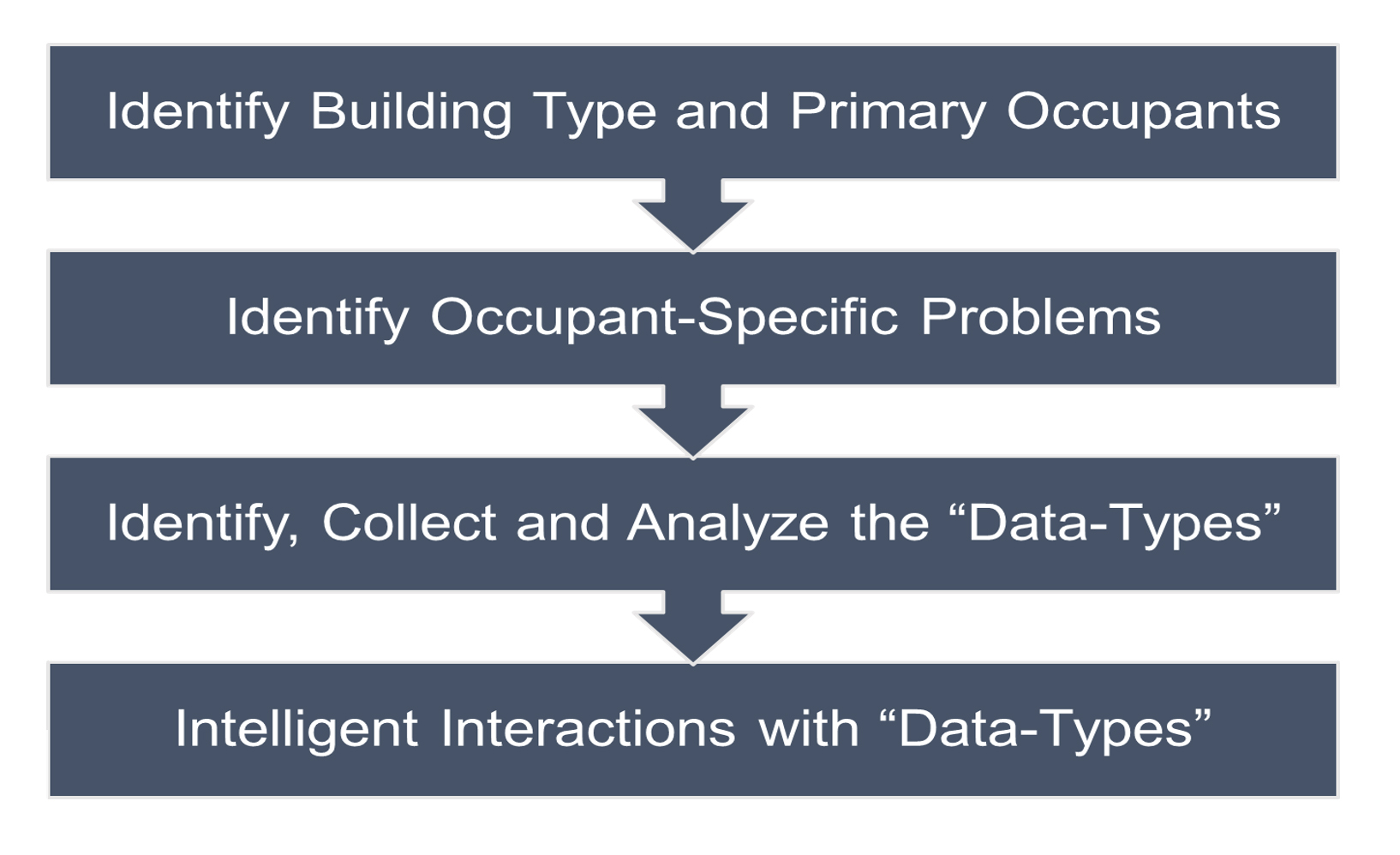 }
\caption{\footnotesize \label{fig7}Application steps of the proposed Framework to specific-purpose buildings}
\end{figure}

\subsubsection{\upshape 5.1 Identify Building Type and Primary Occupants}
The type of the building and its primary occupants should be identified as the first step towards addressing occupant-specific problems in interacting with buildings. This is because the occupant-building relationship is dynamic in nature and can change depending on the purpose of occupancy. We present three scenarios to explain this phenomenon:
1. Different occupants might have different needs from the same building type, e.g., a student might need information about the study resources in academic building whereas a facility manager might need to know about energy efficiency, or air quality of the building.
2. Different occupants might have the same need from the same building type, e.g., both student and facility managers might need a parking space near the academic building.
3. The same occupants might have different needs from the same building type; e.g., one student might need information about the study resources, but another might want to know about the events happening in an academic building. 
Ideally all building occupants should benefit from the improved interaction with buildings, but priority should be given to addressing the needs of primary occupants to offer pleasant occupancy.

\subsubsection{\upshape5.2 Identify and Analyze Occupant-Specific Needs}
Occupant-specific needs can be identified through collecting participatory data from participants and monitoring their behaviors using sensing technologies. Participatory data collection involves conducting semi-structured interviews, focus groups discussions, surveys, etc. where occupants are directly inquired about their needs and problems in a building. Sensing techniques such as thermal cameras can sense if an occupant is showing signs of thermal discomfort. Other occupant concerns such as data privacy and protection when collecting participatory and sensory data should also be acknowledged.

\subsubsection{\upshape5.3 Identify “Data-Types”, and Collect and Analyze the Data}
Once the occupant-specific needs and problems are identified, the next step is to identify various “data-types” that are currently being exchanged or should be potentially exchanged between occupants and building to address the identified occupant-specific needs. “Occupant data-types” can be described as the data needed from the occupants and “building data-types” can be described as the data that has to be collected from the building to address the needs of its occupant. The types of sensory data that are currently being exchanged can be identified by examining the existing infrastructure and BMS of the building. 
A thorough understanding and collection of all the required “data-types” will lead to the development of an “Occupant-Building Data Pool” that gives an estimate of the existing and potential occupant-building relationships (Figure 8). The length of “Interaction Arrow” denotes the efficiency of interaction between the occupants and buildings. The area of “Data Pool Circle” denotes the amount of data shared between the occupants and the building. By increasing the amount of data sharing, the area of the circle will be bigger and consequently, the length of arrow i.e., efficiency of interaction, will increase. The concept of data pool is adapted from the concept of Data Hungry Homes (DHH) \citep{bib36}. DHHs do not simply collect, create, use or transmit data but are hungry for it and need routine data feeding for them to function.

\begin{figure}[ht]
    \centering
    \captionsetup{justification=centering}
\includegraphics[width= 5cm]{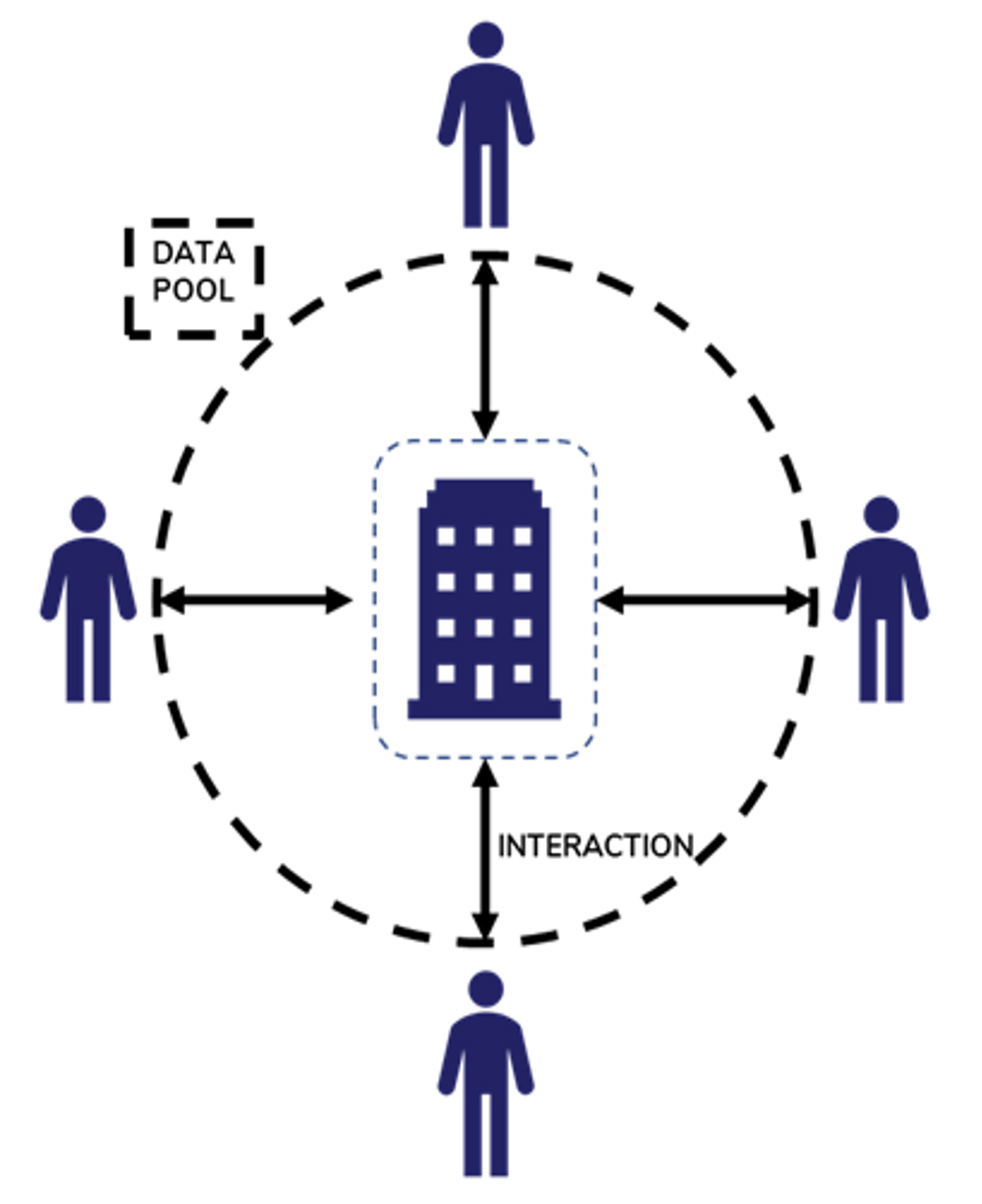 }
\caption{\footnotesize \label{fig8}Occupant-Building Data Pool}
\end{figure}
As mentioned in our proposed vision of HBI, the collection of data for the identified “data-types” can be via sensing and participation for the occupants and via sensing and databases for buildings. Commonly used technologies to collect “occupant data-type” include Infrared (IR) tags, Radio Frequency Identification Devices (RFID) tags, Bluetooth Low Energy (BLE) tags, Ultra-Wide Band (UWB) tags, cameras, mobile phones, smart bands, and interfaces that take their real-time feedback. The sensors that collect “building data-type” include energy meters, temperature sensors, relative humidity sensors (RH), light sensors, sound sensors, smoke sensors, CO\textsubscript{2} and CO sensors, etc. Technologies such as Building Information Modelling (BIM) can be used to create a database for buildings. The selection of technologies for collecting the required data depends on the existing sensing infrastructure of the building. If the existing infrastructure does not support the sensing of a data-type, then the most effective sensing technique for that data-type should be added.

\subsubsection{\upshape5.4 Intelligent Interactions with “Data-Types”}
Don Norman in his book "The Design of Everyday Things"\citep{bib35}, describes how difficult it was for him to operate a building element as simple as the doors. Since modern buildings have more complex elements than just doors, the struggles occupants might face when operating such elements can hinder their experience in the building. To minimize such struggles, building elements, including interfaces, should be designed to maximize building "affordances". The concept of affordances was defined by J. J. Gibson as the information associated with an object about how people could interpret and interact with it \citep{bib35}. Gibson states that affordances are independent of perception, meaning perceptual information is always stored in an object even if it is not perceived by someone. Norman emphasizes the importance of perceivable affordances and uses the term "signifiers" to indicate perceivable affordances. Traditionally, building designers have relied on semiotics to show the necessary information to the users inside a building. However, there are several underutilized and hidden affordances in a building that can be exploited by using advanced interfaces such as Augmented Reality (AR) and smart building-informed artifacts. Building artifacts when equipped with technology can act as interaction channels between the occupants and buildings. As an example, the concept of “smart desks” were developed to address individual indoor environment preferences by using sensors to monitor the environment and ML for learning occupant preference \citep{bib22}\citep{bib37}. Also, a rule-based chat-bot was developed to manage plugged-in appliances through smart plugs in an office environment to involve occupants in the building's energy management and encourage energy savings \citep{bib38}. Another example is the use of programmable robots to control physical objects and reconfigure surrounding environments, also called programmable environments. Researchers have demonstrated the use of such robots to create modular and reconfigurable rooms \citep{bib39}. These rooms have retractable floor tiles that can rise from the floor to provide places to sit and shape-changing walls to provide privacy to the occupants when needed.

\subsection{6. Case study: Enhancing learning experience of students in Goodwin Hall}
In this section, a step-by-step application of the proposed framework to enhance the learning experience of the students in Goodwin Hall, as a representative technology-enhanced building at Virginia Tech, is presented. 

\subsubsection{\upshape6.1 Goodwin Hall Building Type and its Primary Occupants}
Goodwin Hall is an academic building at Virginia Tech. It is the flagship smart building which houses 40 instructional and research labs, 8 classrooms, an auditorium, and 150 offices for several engineering departments. Goodwin Hall has around 240 accelerometers attached to 136 sensor mounts throughout the building’s ceilings that can measure all vibrations made inside the building. It is also equipped with temperature and CO2 sensors to regulate the indoor climate. Primary occupants of Goodwin Hall include undergraduate and graduate students whereas the secondary occupants are faculty members and administrative staff. The primary purpose of any academic building like Goodwin Hall should be to provide an adequate and efficient learning experience for the students.

\subsubsection{\upshape6.2 Occupant-Specific Needs in Goodwin Hall }
In order to identify the students' needs and preferences while occupying the building, we conducted focus groups and semi-structured interviews. The primary occupants were defined as students who spend more than 10 hours per week in Goodwin Hall. Qualitative study and analysis were conducted to identify important categories. These categories are grouped using concept mapping to identify various themes and relationships among categories, codes, and subcodes.

\paragraph{\emph{6.2.1 Qualitative Data Collection}}\hfill \break
Focus groups and semi-structured interviews were conducted among 12 undergraduate and 7 graduate students. The participating students spent at least 10 hours per week in Goodwin Hall. The questions in the focus groups and semi-structured interviews asked participants about their positive and negative experiences in the building, potential applications, and advantages of building IUIs with respect to improving the learning experience, student's preferred interaction modalities, and potential data privacy concerns. All focus groups and semi-structured interviews were audio-recorded with consent from participating students. 

\paragraph{\emph{6.2.2 Qualitative Data Analysis}}\hfill \break
The audio data was transcribed to text and qualitative coding (NVivo and Focused tools) was used to identify the important categories, codes, and subcodes in the raw text data. Initially, audio to text transcription was done using web-based services that resulted in a total of 12 transcribed documents. These documents were manually checked to remove possible errors in transcription. NVivo coding was used to assign labels to the data collected from the focus groups and semi-structured interviews. It uses words or short phrases from the student’s sentences in the transcribed data as codes. Focused coding was further used to categorize most frequent or significant codes and subcodes to develop categories. Qualitative coding was conducted using MAXQDA software. Two iterations were conducted to validate the reliability of the coding that resulted in a total of 580 NVivo codes. In the first iteration, 228 NVivo codes were identified and in the second iteration, the number of NVivo codes was 352. Both iterations were performed by the same researcher in a gap of 2 weeks to account for unnecessary bias in qualitative coding. Intra-coder reliability tests were performed on the NVivo codes from the two iterations to check their validity. Percent agreement and Cohen’s Kappa statistic were used to check the intra-coder reliability of the NVivo codes. The purpose of intra-coder reliability is to improve the quality of the identified NVivo codes. The result of the intra-coder reliability test is shown in Figure 9.
\begin{figure}[ht]
    \centering
    \captionsetup{justification=centering}
\includegraphics[width= 12cm]{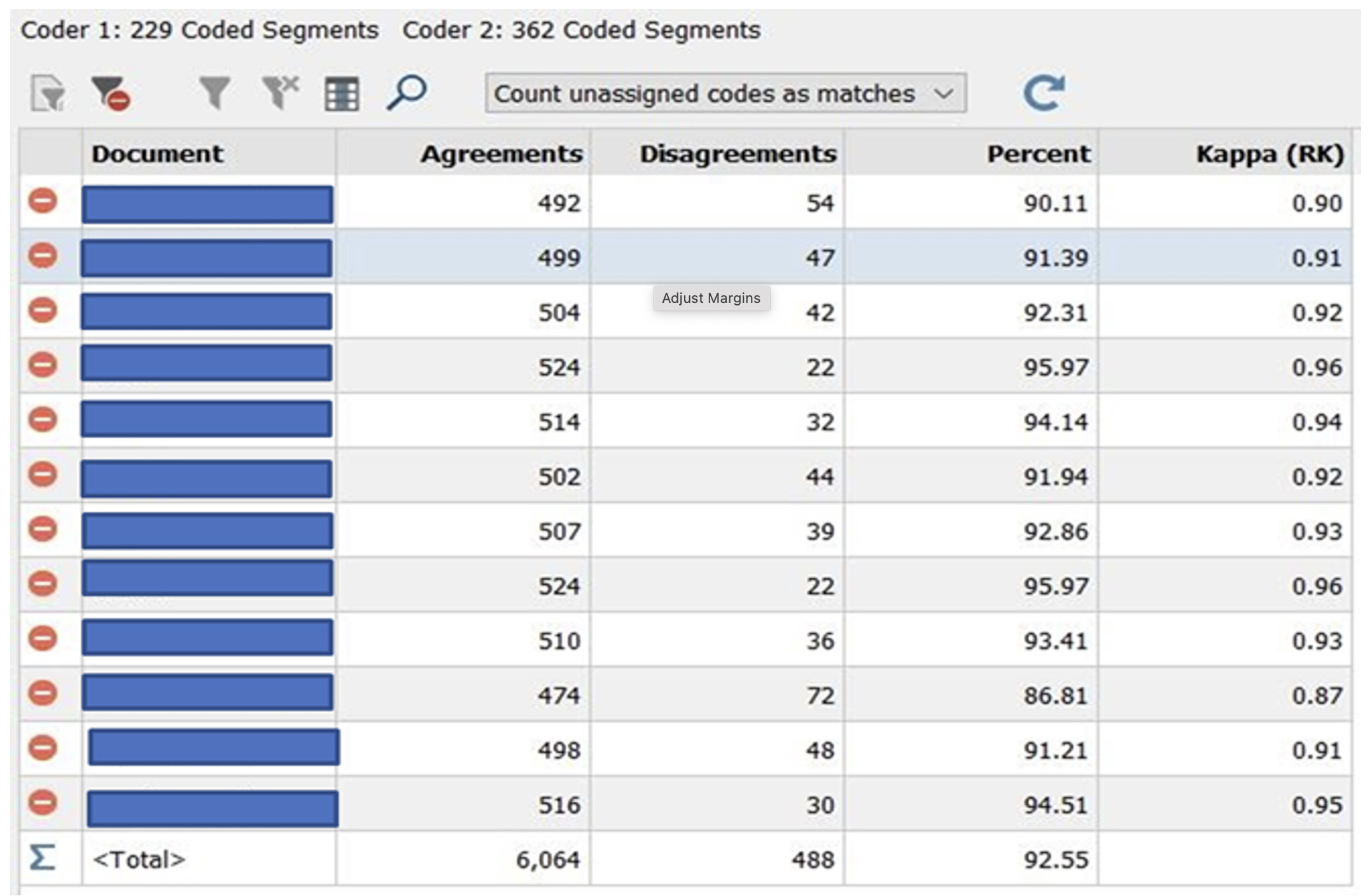 }
\caption{\footnotesize \label{fig9}Percent Agreement and Cohen's Kappa Statistic for Intra-Coder Reliability test}
\end{figure}

\begin{table}
\caption{Frequency and Weight of categories, codes and subcodes identified from Focused Coding} \label{exp_dataset1}
\begin{tabular*}{\hsize}{ | p{4.15cm} | p{0.8cm}| p{1.2cm} | p{1.6cm}| p{1.37cm}|  }

\hline
 \large\textbf {Categories and codes} & \large\textbf{Freq} & \large\textbf{Code weight} & \large\textbf{Subcode weight} & \large\textbf{Sub-subcode weight} \\
 \hline
 \textbf{DATA PROTECTION} & 23 & 8 & & \\
 \hline DATA SHARING & 5 & 8 & 3 & \\
 \hline OTHERS& 17 &  & 6 & \\
 \hline SCHEDULE &  & 8 &  & \\
 \hline \textbf {INTERACTION MODALITY} & 6 & 8 & 5 & \\
 \hline OTHERS & 25 &  & 6 & \\
 \hline TOUCH & 6 &  & 6 & \\
 \hline AUDIO & 18 &  & 8 & \\
 \hline VISION & 21 & & & \\
 \hline \textbf {OCCUPANT-SPECIFIC PROBLEM} & 187 & 10 & 5 & \\
 \hline OTHERS & 17 &  & 5 & \\
 \hline EVENT NOTIFICATION & 3 & & 5 & \\
 \hline FINDING PROFESSORS & 6 & & 8 & \\
 \hline FINDING PARKING SPACES & & 8 & 3 & \\
 \hline SAFETY & 6 &  & 6 &  \\
 \hline STUDY RESOURCES & 33 &  & 9 &\\
 \hline others & 2 &  &   & 3 \\
 \hline setting reminders & 3 & & & 3 \\
 \hline smart furniture & 10 & & & 8 \\
 \hline power outlets & 4 & & & 6 \\
 \hline smart boards & 14 & & & 9 \\
 \hline FINDING STUDY SPACES & 20 &  & 8 & \\
 \hline STRESS MANAGEMENT & 18 & & 8 & \\
 \hline COMFORT & 61 &  & 10 & \\
 \hline others & 4 &  &   & 3 \\
 \hline food & 4 &  &   & 5 \\
 \hline auto doors & 3 &  &   & 5 \\
 \hline visual & 20 &  &   & 9 \\
 \hline audio & 21 &  &   & 9 \\
 \hline temperature & 9 &  &   & 9 \\
 \hline INDOOR NAVIGATION & 14 &  & 7  &   \\
 \hline TRAFFIC MANAGEMENT & 5 &  & 5 &   \\
\hline
\end{tabular*}
\end{table}

Focused coding was performed on the second iteration to group significant NVivo codes into categories, codes and subcodes. 301 of the 352 NVivo codes were considered for focused coding and the rest were discarded because they were insignificant. All the categories, codes and subcodes were assigned a weight based on their frequency of occurrence and the emphasis given by the participants in the focus groups and semi-structured interviews (Table 1). Concept maps were also created to explore relationships among the identified codes and subcodes to assess its impact on the learning experience of students in the building (Figure 10) 
\begin{figure}[ht]
    \centering
    \captionsetup{justification=centering}
\includegraphics[width= 10cm]{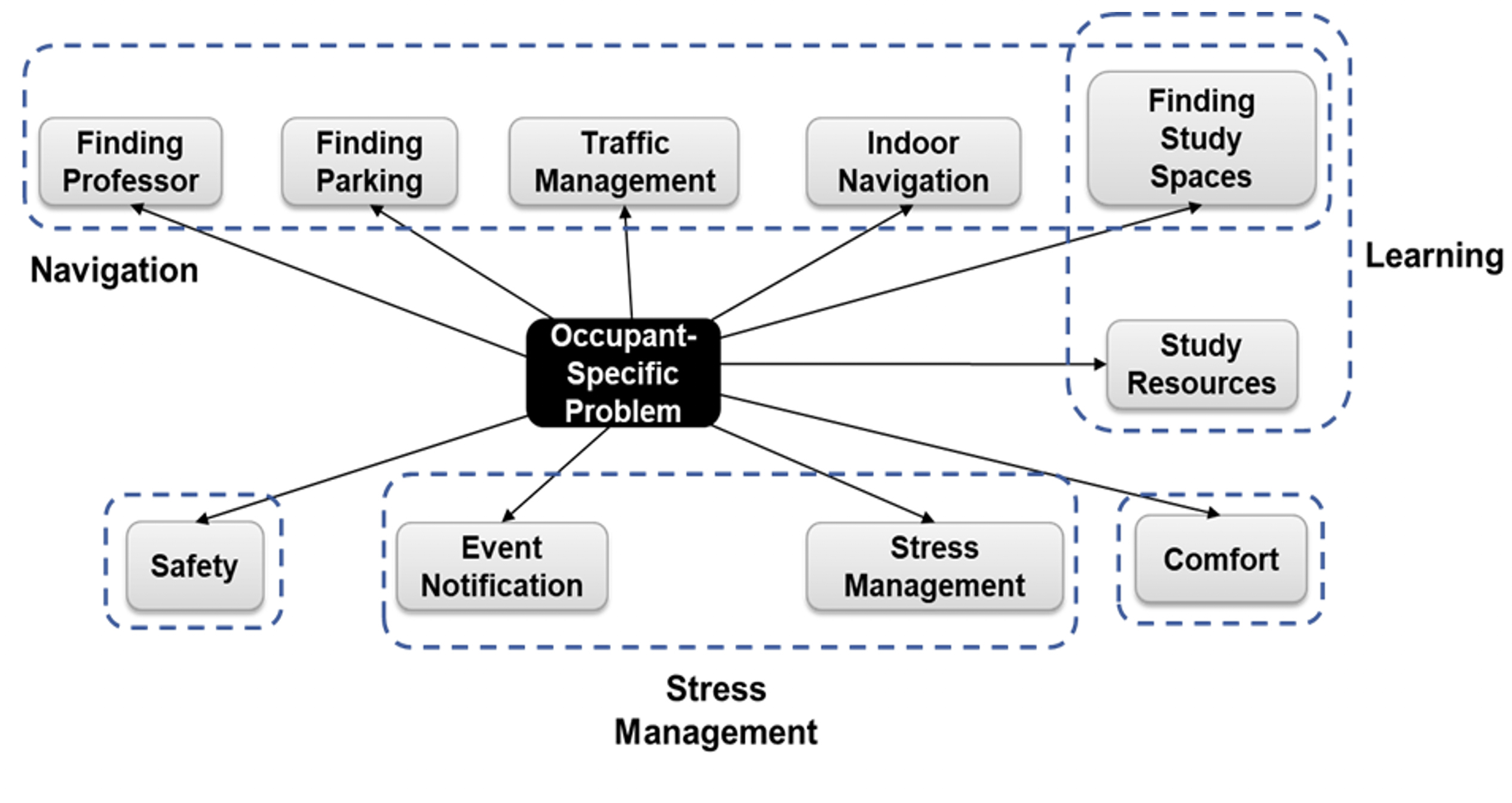}
\caption{\footnotesize \label{fig10}Concept Map showing code relationships of category “Occupant-Specific Problem”}
\end{figure}
\paragraph{\emph{6.2.3 Results of Qualitative Data Analysis}}\hfill \break
The key themes that emerge from the concept mapping exercise are “Navigation”, “Learning”, “Comfort”, “Stress Management” and “Safety”.  The theme “Navigation” includes the codes “finding professor”, “finding parking”, “traffic management”, “indoor navigation”, and “finding study spaces”. “Learning” has codes “finding study spaces” and “study resources”. “Comfort” is concerned with “audio”, “visual” and “thermal” comfort levels of the occupants. “Stress management” includes “sending notification” about the “events” happening in the building. “Safety” includes the protection of occupants in case of an emergency such as fire alarm and shooting event inside the building. A summary of the themes with their cumulative weights is shown in Table 2. The theme with the highest cumulative weight is “Comfort”. “Navigation” and “Learning” also have significant weights assigned to them.

\begin{table*}[ht]

\caption{Themes and Cumulative Weights in category “Occupant-Specific Problem”} \label{exp_dataset2}

\begin{tabular*}{\hsize}{| p{4.15cm} | p{4.5cm}| p{1.5cm} | p{2.7cm}| }
\hline
 \large\textbf {Theme} & \large\textbf{Codes} & \large\textbf{Weight} & \large\textbf{Cumulative Weight} \\
\hline
 \multirow {5}{*}{NAVIGATION} & FINDING PROFESSORS & 6 &   \\
              \cline{2-3}    &   FINDING PARKING SPACES &  4 &    \\
               \cline{2-3}  &   INDOOR NAVIGATION &  14 &   49 \\
                 \cline{2-3} &  TRAFFIC MANAGEMENT &  5 &   \\
                 \cline{2-3} &  FINDING STUDY SPACES&   20 &    \\
 \hline \multirow {2}{*}{LEARNING} & STUDY RESOURCES& 33 & 53 \\
                             & FINDING STUDY SPACES & 20 &   \\
 \hline COMFORT & COMFORT & 61 & 61 \\
 \hline \multirow{2}{*} {STRESS MANAGEMENT} & STRESS MANAGEMENT  & 18 & 21 \\
                                      & EVENT NOTIFICATION & 3 &  \\
 \hline SAFETY & SAFETY & 6 & 6 \\

\hline
\end{tabular*}
\end{table*}
The codes with indirect impact on students’ learning experience are “comfort”, “finding professor”, “stress management”, “navigation”, “parking”, “traffic management”, “safety”, and “event notification”. Figure 11 shows a structured classification of these codes along with their relative importance percentages. The most important code observed in Figure 11 is “comfort” and although it does not directly contribute to the learning experience, it has a significant indirect impact on enhancing the learning experience by providing students with a comfortable environment. “Comfort” was a primary concern for the graduate students who spent more time in Goodwin Hall. They complained about the lack of windows in their labs and how it makes them feel disconnected from the outside world and sometimes even claustrophobic. Lack of “audio comfort” was a problem for both the undergraduate and graduate students because the undergrads preferred “study spaces” that are less noisy, and the grads complained about the noises that disturbed them while they were working in their labs. “Thermal comfort” was also important for all the students, they required more control of and contribution to regulating the indoor temperature. “Stress management” was a concern for the graduate students because they spend a lot of time in the building and need something to help them relax in between their studies. A requirement and a possible solution for “stress management” is “event notification”, to inform the students about the events happening in building so they can attend in their free time. “Finding professors” was also a concern for the graduate students because most of them are working as graduate assistants with a professor and sometimes face problems in locating their respective professors. “Traffic management” refers to “finding parking” around receiving information regarding parking availability at different times. providing “safety measures” against possible emergencies such as fire and intruder was highlighted as a requirement by the students. They were also concerned about their data privacy and security but were willing to share their location and schedule data as long as the data does not have any identifiers and remains anonymous.

\paragraph{\emph{6.2.4 Discussions on Qualitative Data Analysis}}\hfill \break
This section presents qualitative evidence that validates the hypothesis of this research, i.e., occupant-specific needs in academic buildings can have a negative impact on the learning experience of students. Based on the obtained results, all the identified factors directly or indirectly impact the learning experience of the students in an academic building. The theme “Learning” has a direct impact on the learning experiences whereas other identified factors make an indirect impact. The codes with direct impact are: “finding study spaces” and “study resources”. Many undergrad students highlighted that even though Goodwin Hall has a lot of open spaces, it was difficult for them to find an appropriate place to study in the building. Finding study spaces is a common problem that many undergraduate and graduate students usually face in academic buildings and libraries. Students also expressed the difficulties they faced in finding study resources such as proper furniture, power outlets, and whiteboards in the buildings. 
\begin{figure}[ht]
    \centering
    \captionsetup{justification=centering}
\includegraphics[width= 12cm]{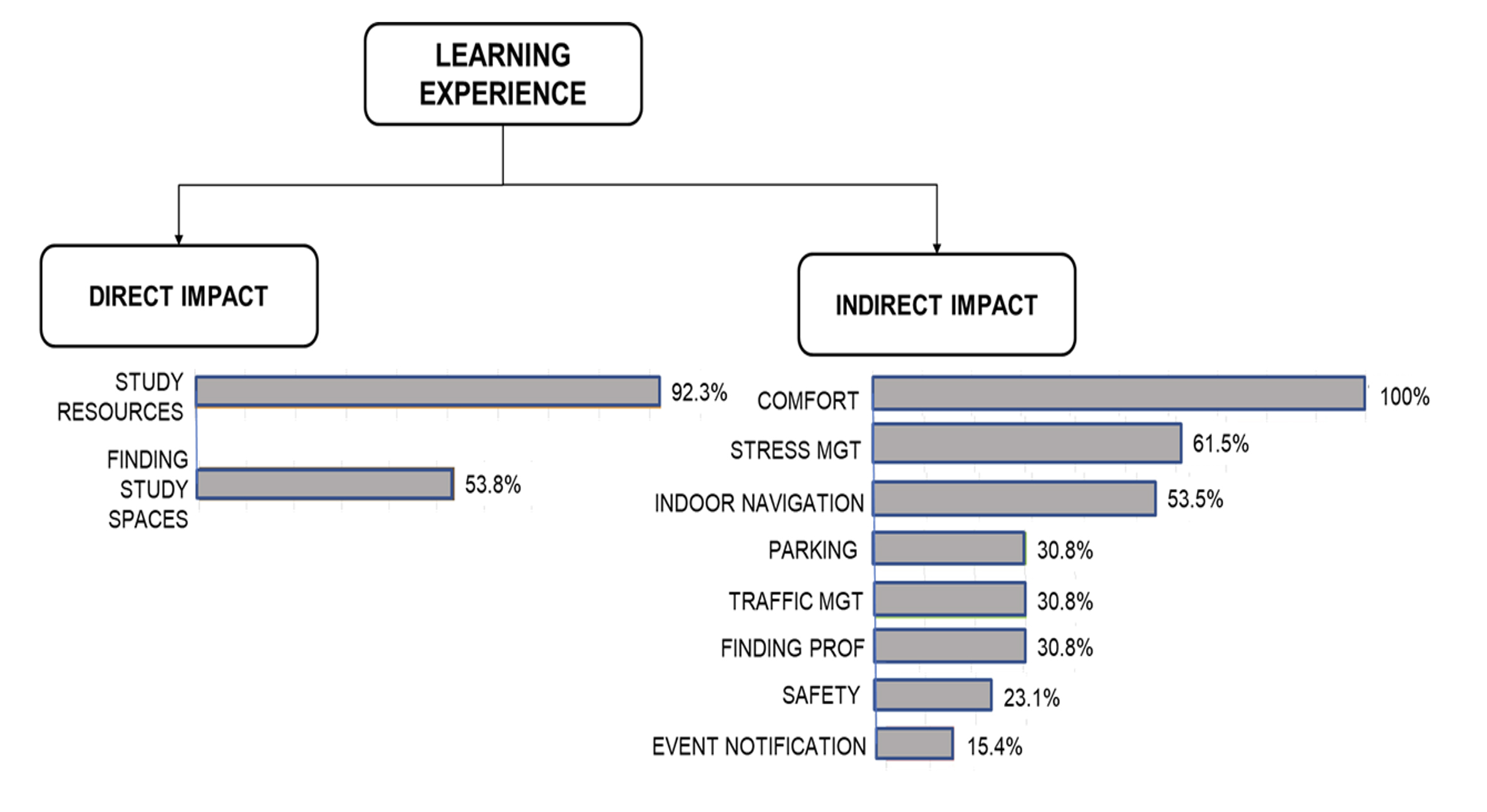}
\caption{\footnotesize \label{fig11}Relative Importance Percentages of identified codes on Learning Experience}
\end{figure}

\subsubsection{\upshape 6.3 Identify and Collect “Data-Types” from Goodwin Hall}
Regarding Goodwin Hall, the “occupant data-types” and “building data-types” identified from the theme “navigation” and the respective sensing and interaction strategies are shown in Table 3. It can be observed that most of the “building data-types” are concerned with location, that can make use of Goodwin Hall’s accelerometers. A downside of accelerometers is that they can only determine the location of moving objects and not stationary objects. For the location of stationary objects, sensors such as RFID, BLE, and Wi-Fi can be used. 
\begin{table*}[ht]
\caption{\label{exp_dataset3}Data-Types Collection and Interaction Strategies for theme “Navigation”}
\begin{tabular*}{\hsize}{| p{1.2cm} | p{1.2cm}| p{2cm} | p{1.2cm}| p{1.2cm} | p{1.9cm}| p{1.2cm} | p{1.2cm}|p{1.5cm}|}
\hline
\textbf {Codes} &\textbf  {Building data type} & \textbf {Existing collection strategy} & \textbf {Potential collection strategy} &\textbf  {Occupant data type} & \textbf {Existing collection strategy} & \textbf {Potential collection strategy} & \textbf {Existing interaction strategy} \\
\hline
 \multirow{2}{*}{}Finding Professors & Professor Location  & Accelerometer & RFID, BLE, Wi-Fi & Student Location & Accelerometer  & RFID, BLE, Wi-Fi & None \\
                               \cline{2-8}       & Professor Schedule &None & Professor Calendar  & Student Location & None & Student Calendar &None  \\
                                     \hline
 Finding Parking Space &Empty Parking Space &None & Proximity Sensor  & Student Location & Accelerometer & GPS &None \\
 \hline Finding Study Space & Study Space Location &None & RFID, BLE, Wi-Fi & Student Location & Accelerometer & RFID, BLE, Wi-Fi  &None  \\
 \hline Indoor Navigation & Room Location &Building Map, Building 3D Model  & NA & Student Location &Accelerometer & RFID, BLE, Wi-Fi  &None \\
 \hline Traffic Management & Occupant Location &Accelerator Meter& RFID, BLE, Wi-Fi & Student Location & Accelerometer & RFID, BLE, Wi-Fi  &Screens  \\
\hline
\end{tabular*}
\end{table*}
\begin{table*}[ht]
\caption{\label{exp_dataset4}Data Collection and Interaction Strategies for theme “Learning”}
\begin{tabular*}{\hsize}{| p{0.9cm} | p{1.2cm}| p{1.2cm} | p{1cm}| p{1.3cm} | p{1.2cm}| p{1.4cm} | p{1.2cm}|p{1.27cm}|}
\hline
\textbf {Codes} & \textbf {Subcodes} & \textbf  {Building data type} & \textbf {Existing collection strategy} & \textbf {Potential collection strategy} &\textbf  {Occupant data type} & \textbf {Existing collection strategy} & \textbf {Potential collection strategy} & \textbf {Existing interaction strategy} \\
\hline
  Study Resources & setting reminders & NA & NA & NA & Student Schedule &None & Student Calendar & None \\
 \hline  & furniture & Furniture Location & None & RFID, BLE, Wi-Fi & Student Location & Accelero -meter & RFID, BLE, Wi-Fi & None\\
\hline   & power outlets & Power Outlet Location & NA & Building Map, Building 3D Model & Student Location & Accelero -meter & RFID, BLE, Wi-Fi & None \\
\hline  & white boards & White board Location & None & RFID, BLE, Wi-Fi & Student Location & Accelero -meter & RFID, BLE, Wi-Fi & None \\
\hline Finding Study Space &   & Study Space Location & None & RFID, BLE, Wi-Fi & Student Location & Accelero -meter & RFID, BLE, Wi-Fi & None \\
\hline
\end{tabular*}
\end{table*}
In addition to location, student comfort preferences also contribute to the “student data-types”. “Building data-types” for “Comfort” include luminance, noise level, and temperature and these are measured with light sensors, microphones, and temperature sensors, respectively. “Occupant data-types” for “Comfort” include location and subjective feedback through surveys. These along with sensing techniques and interaction strategies for “Comfort” are shown in Table 5. 
\begin{table*}[ht]
\caption{\label{exp_dataset5}Data Collection and Interaction Strategies for theme “Comfort”}
\begin{tabular*}{\hsize}{| p{1.05cm} | p{1.3cm}| p{1.2cm} | p{1.2cm}| p{1.2cm} | p{1.2cm}| p{1.2cm} | p{1.2cm}|p{1.15cm}|}
\hline
\textbf {Codes} & \textbf {Subcodes} & \textbf  {Building data type} & \textbf {Existing collection strategy} & \textbf {Potential collection strategy} &\textbf  {Occupant data type} & \textbf {Existing collection strategy} & \textbf {Potential collection strategy} & \textbf {Existing interaction strategy} \\
\hline
 Comfort &   &  &   &   &   &   &  &   \\
 \hline & food &ABP Menu & NONE & ABP Website & Student Preferences  & None &Survey & None \\
\hline  & auto doors &NA & NA & NA  & Student Location & Accelero -meter &RFID, BLE, Wi-Fi & None \\
\hline  & visual &Lumin -ance & NA & Light Sensor & Student Preferences  & None &Survey & None \\
\hline  & audio &Noise Level & NA & Micro -phone & Student Preferences  & None &Survey & None \\
\hline  & temper -ature &Temper -ature & NA & NA & Student Preferences  & None &Survey & Thermo -stat \\
\hline
\end{tabular*}
\end{table*}

The “occupant data-types” and “building data-types” along with their respective sensing techniques and interaction strategies for the theme “Stress management” and “Safety” are shown in Table 6. It can be observed in Table 6 that location and schedule constitute to be important data-types for “Stress management” and “Safety” as well.  
\begin{table*}[ht]
\caption{\label{exp_dataset6}Data-Types Collection and Interaction Strategies for theme “Stress Management” and “Safety”}
\begin{tabular*}{\hsize}{| p{0.9cm} | p{1.2cm}| p{1.2cm} | p{1.2cm}| p{1.3cm} | p{1.2cm}| p{1.2cm} | p{1.2cm}|p{1.27cm}|}
\hline
\textbf {Codes} & \textbf {Subcodes} & \textbf  {Building data type} & \textbf {Existing collection strategy} & \textbf {Potential collection strategy} &\textbf  {Occupant data type} & \textbf {Existing collection strategy} & \textbf {Potential collection strategy} & \textbf {Existing interaction strategy} \\
\hline
 Event Notification & XX &Event Schedule & None & VT Event Planning Calendar  &Student Schedule & None &Student Calendar & Screens \\
 \hline Stress Management  & XX &NA &NA & NA & Stress Level & None &Face recognition, Survey & None \\
 \hline \multirow {2}{*} {Safety} & fire &Fire Location & Smoke Detector& Flame Detector & Student Location & Accelero -meter &RFID, BLE, Wi-Fi & Fire Alarms \\
        \cline{2-9} & intruder &Intruder Location & Accelero -meter & RFID, BLE, Wi-Fi& Student Location & Accelero -meter &RFID, BLE, Wi-Fi & None \\
 
\hline
\end{tabular*}
\end{table*}
\subsubsection{\upshape6.4 Potential Interaction Strategies to address Occupant needs in Goodwin Hall}
In this section, we discuss some conceptual design ideas for the smart building interaction systems that can address the above-mentioned occupant-building specific needs in our targeted building.
\paragraph{\emph{6.4.1 Augment Building Affordances using AR}}\hfill \break
The existing interaction strategies in Goodwin Hall include screens located at the main entrance to show expected occupancy, thermostats to show the current temperature, and fire alarms to warn in case of a fire emergency. The interaction modalities preferred by the students in descending order of preference include vision, audio, touch, and gestures. As discussed previously, after identifying data-types and their sources, and collecting the data from building and occupants, the collected data should be processed, and the extracted information should be conveyed to the occupants. To this end, the building should be equipped with interaction systems to exchange the information with occupants who need it. The concept of affordances can be used when using the building as a medium of exchange. Traditionally, building designers have relied on semiotics to show necessary information to the users inside a building, but there are several underutilized affordances in a building which can be exploited by using Augmented Reality (AR). The capability of AR to superimpose valuable information processed from the data types on top of physical objects and its ability to interact with the occupant using various modalities makes it a good fit for the smart building interaction systems. Other information exchange means can include basic artifacts such as furniture, screens, etc. and smart artifacts such as chat-bots, robots, etc.

\paragraph{\emph{6.4.2 Context-aware ubiquitous analytics through smart learning environments}}\hfill \break
Learning Analytics (LA) is defined as "the collection, analysis, use, and appropriate dissemination of student-generated, actionable data to create appropriate cognitive, administrative, and effective support for learners"\citep{bib41}. LA-based strategy includes several advantages for both students and instructors. this strategy enables instructors to identify students' learning patterns and profiles, monitor students' trajectory and potential problems, promote their teaching strategies and educational resources, etc. \citep{bib42}. LA's benefits for students include monitoring their learning process and comparing their performance with other students, identifying their goals, taking control of their learning process, etc. Current LA tools such as Moodle and Blackboard do not include contextual data of students such as locations and times \citep{bib43}. Ubiquitous learning analytics and contextual learning analytics have emerged as LA's sub-domains in the context of a smart learning environment. These methods aim to "understand and optimize learning and the environments in which learning occurs through the measurement, collection, analysis, and reporting of data about learners and their contexts"\citep{bib44}. 
Regarding the problems identified in this research for the theme "Learning", i.e., "studying resources" and "finding study spaces", smart pedagogy which is "a science that is constantly evolving and looking for ways to teach better and to scaffold students in the process of knowledge building"\citep{bib45} can present viable solution to enhance students' learning experience in an academic building. For instance, for addressing "stress management" and "finding professors" that are demonstrated as the categories with a significant indirect impact on students' learning experience, assessing students' psychological conditions, such as their emotions, motivation, and stress levels, can be collected and used through building to enhance their learning environment. In this sense, providing personalized services to each student, such as personalized teaching and learning services and psychological services, can be implemented. Personalized learning includes aspects such as personalized goal planning, personalized course profile and syllabus, personalized digital contents, customized grading system, flexible, collaborative works, and formative assessment [46]. These can be provided to the students when using different spaces, and depending on their shared schedules. Additionally, efficient collaboration and interaction between students and with their instructors are other factors that affect the quality of learning and teaching environments. 
Considering the cases mentioned above, a context-aware ubiquitous smart learning and teaching strategy can be used to augment academic buildings and address the identified problems and improve learning quality. Such context-aware ubiquitous learning systems can use mobile devices, wireless communication, and sensing technologies such as RFID-tags, QR-codes, and GPS is to collect required information from occupants and provide personalized or adaptive learning support based on learners' preferences, learning skills, and contexts of learners [47]. 

\subsection{7. Conclusion and Future Works}
This paper compiles existing research on “Human-Building Interaction (HBI)” to identify gaps in HBI theories and its applications. We propose a novel vision to develop a holistic HBI framework that integrates the concepts of HCI and AmI with HBI to increase the intelligence of building and their interaction with occupants through create intelligent user interfaces (IUIs) that facilitate real-time, bidirectional, and multi-modal interactions between occupants and buildings. Additionally, it was observed that the current HBI approaches are general and lack applications for addressing the needs of occupants in specific-purpose building including academic buildings. Therefore, we apply the proposed HBI system to study the needs of primary occupants, i.e., students, in academic buildings and suggested strategies to address the identified needs. We used a human-centered approach to identify the occupant-specific problems in Goodwin Hall (flagship engineering building at Virginia Tech) and evaluate its impact on the learning experience of students. Focus groups and semi-structured interviews were conducted among 12 undergraduate and 7 graduate students to identify their problems. Analysis of the qualitative data performed by using qualitative coding and concept mapping revealed that the key occupant-specific problem themes are related to “Navigation”, “Learning”, “Comfort”, “Stress Management” and “Safety”. Qualitative data analysis also revealed that the code “study resources” has highest direct impact and code “comfort” has highest indirect impact on the learning experience of students in Goodwin Hall. The existing and potential “data-types”, existing and potential sensing techniques and existing interfaces are studied to create a “data pool” for Goodwin Hall. There are accelerometers, temperature sensors, humidity sensors and CO2 sensors in Goodwin Hall but additional sensors to measure occupancy, comfort, stress, and safety are needed. Based on the identified themes from qualitative data analysis, their respective “data-types” and existing interaction strategies in Goodwin Hall, the features of the HBI system for Goodwin Hall should be such that it makes indoor navigation easier, provides essential study resources, helps in maintaining indoor environmental quality (IEQ), and ensures a stress-free and safe social atmosphere. Towards the end, we discuss various interaction principles including building affordances using AR and context-aware ubiquitous analytics-based strategy to design the IUI for Goodwin Hall. The former aims to superimpose useful information on the top of physical objects to interact with occupants and the latter suggests ubiquitous personalized services encompassing teaching/learning services and psychological services based on occupants' contexts. In our future work, we will establish a HBI system for Goodwin Hall and perform usability studies on the HBI system to measure its impact in mitigating the occupant-specific problems and in enhancing the learning experience of students in Goodwin Hall.




\end{document}